\documentclass[aps,prx,preprint,showpacs,nofootinbib,superscriptaddress]{revtex4-1}
\usepackage{latexsym}
\usepackage{amssymb}
\usepackage{graphicx}
\usepackage{amsmath}
\usepackage{bm}
\usepackage[colorlinks,
          linkcolor=black,
            citecolor=black,
            urlcolor=blue
           ]{hyperref}
\usepackage{verbatim}
\usepackage{mathrsfs}
\usepackage{extarrows}
\usepackage{comment}
\usepackage{mathtools,slashed}

\begin{document}

\title{Nonlinear parameter-gauge coupling approach to generalization of generalized Thouless pumps and $-1$-form anomaly}

\author{Yuan Yao}
\email{smartyao@issp.u-tokyo.ac.jp}
\affiliation{Institute for Solid State Physics, University of Tokyo, Kashiwa, Chiba 277-8581, Japan}

\begin{abstract}
We study the nontrivial topology of the parameter space of general $U(1)$-symmetric fermionic non-degenerately gapped system and its consequences on the transport properties in arbitrary dimensions. 
By a nonlinear parameter-gauge topological response theory, 
we find that such nontrivial topology can impose quantization constraints on the charge transport in the presence of background fluxes or, more generally, instantons in general dimensions
and our result generalizes the Thouless pump and its higher dimensional generalizations. 
We also show that these nontrivial transport properties are related to an unconventional quantum anomaly, which generalizes $-1$-form anomalies. 
This anomaly imposes non-perturbative ingappabilities of various types of spatial interfaces or time-dependent system evolution.

\end{abstract}
\maketitle
\tableofcontents
\newpage

\section{Introduction}
Transportation properties of quantum matter at zero temperature play essential roles in manifesting nontrivial topology of ground-state wavefunction, e.g. the Nobel-prize winning result Thouless-Kohmoto-Nightingale-Nijs formula~\cite{Thouless:1982aa} relates the integer quantum Hall conductance with the Chern number of filled electronic bands. 
Later, the topological nature of an adiabatic cycling of gapped one-dimensional $U(1)$-charge symmetric system was investigated by D. Thouless through a quantized charge transport --- the Thouless charge pump~\cite{Thouless:1983,Niu:1986aa}. 
It is stated that, in a gapped free-electronic system with a unique ground state, the total net charge flowing across a fixed section of a one-dimensional ring within one period is an integer. 

The Thouless charge pump can be also seen in a familiar (bosonic though) interacting system, e.g. a spin-$1$ antiferromagnetic (AFM) Heisenberg chain: 
\begin{eqnarray}
\label{afm}
H_\text{AFM}=\sum_i\vec{S}_i\cdot\vec{S}_{i+1}, 
\end{eqnarray}
which is defined on a spatial ring with a periodic boundary condition and possesses a nonzero gap above the unique ground state~\cite{Haldane:1983aa,Affleck:1989aa}. 
This system has a full $SO(3)$ spin-rotation symmetry, but let us simply focus on the $U(1)_z$ subgroup which is the rotation symmetry along $z$-axis.
Let us impose an external magnetic field in the $z$-direction which is significantly large only in a finite regime of the chain. 
The external field still respects $U(1)_z$ symmetry and, we can switch on other necessary local $U(1)_z$-symmetric interactions where the magnetic field starts vanishing so that the energy spectrum is eventually still gapped with a unique ground state.
Thus the ground state can potentially have a nonzero $S_z$ value (which is the $U(1)_z$-charge) magnetized by the magnetic field and such a $U(1)_z$ charge is quantized trivially. 
In the low-energy limit, we can assume that the system, which has a single finite-energy state asymptotically, is topological and conformal,
e.g. it possesses a Lorentz invariance and an S-invariance --- we can exchange the (one-dimensional) space and the time. 
\begin{figure}[h]
\begin{center}
\includegraphics[width=8.5cm,pagebox=cropbox,clip]{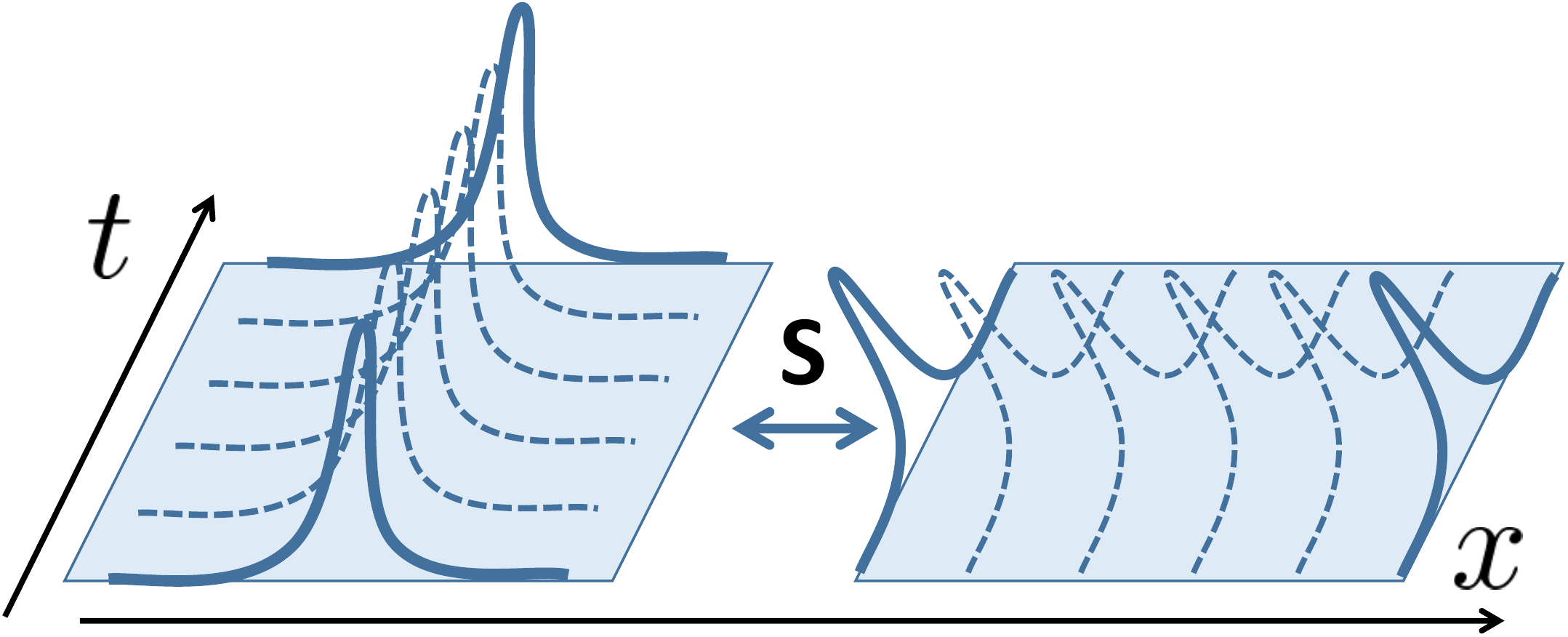}
\caption{The static non-uniform magnetic field in the space-like ``pump'' or jump (left) is S-transformed to a cycling (right). }
\label{S_transformation}
\end{center}
\end{figure}
Let us S-transform the spacetime as in FIG.~(\ref{S_transformation}), 
and then the former magnetism in the presence of the static magnetic field is interpreted as the total spin flowing across a site along the new spatial direction within the new temporal period, illustrated by
\begin{eqnarray}
\int dt J_t(t,x)\leftrightarrow\int dx J_x(t,x)\text{ as }t\leftrightarrow x, 
\end{eqnarray}
where $J_{t,x}$ is the current coupled to the gauge field $A_{t,x}$. 
Such a net spin flow is quantized by the original space-like viewpoint, 
by which we can also see this net flow is robust against perturbation. 
It is because, if we want to smoothly change the magnetism in the former picture,
it is inevitable to close the many-body gap so that the $S_z$-eigenvalue of the ground state can be changed. 
Such a gap closing formally results in large correlation lengths, respectively, in spatial and temporal directions, in two pictures transformed to each other by S-transformation, which contradicts with the correlation length vanishing in the low-energy limit of the later viewpoint. 

The argument above implies that a nontrivial Thouless pump reflects a nontrivial topology of the parameter space of the gapped lattice model~\cite{Kapustin:2020aa,Kapustin:2020ab,Hsin:2020aa} that we will see later. 
For instance, the gapless point in the full parameter space can obstruct the contractibility of the gapped-system parameter space. 
Such gapless defects and their stabilities have been investigated in band topological insulators and superconductors~\cite{Teo:2010aa}. 
Recently, the generalized Thouless pump has been proposed in higher $d$-dimensional space by a linear gauge-field coupling with a $d$-dimensional lattice parameters with $U(1)$ symmetry, and 
Wess-Zumino-Witten terms as generalized Berry curvatures~\cite{Abanov:2000aa,Freed:2008aa} are applied to describe various adiabatic phases without symmetries~\cite{Kapustin:2020aa,Kapustin:2020ab,Hsin:2020aa}. 

\begin{figure}[h]
\begin{center}
\includegraphics[width=6.5cm,pagebox=cropbox,clip]{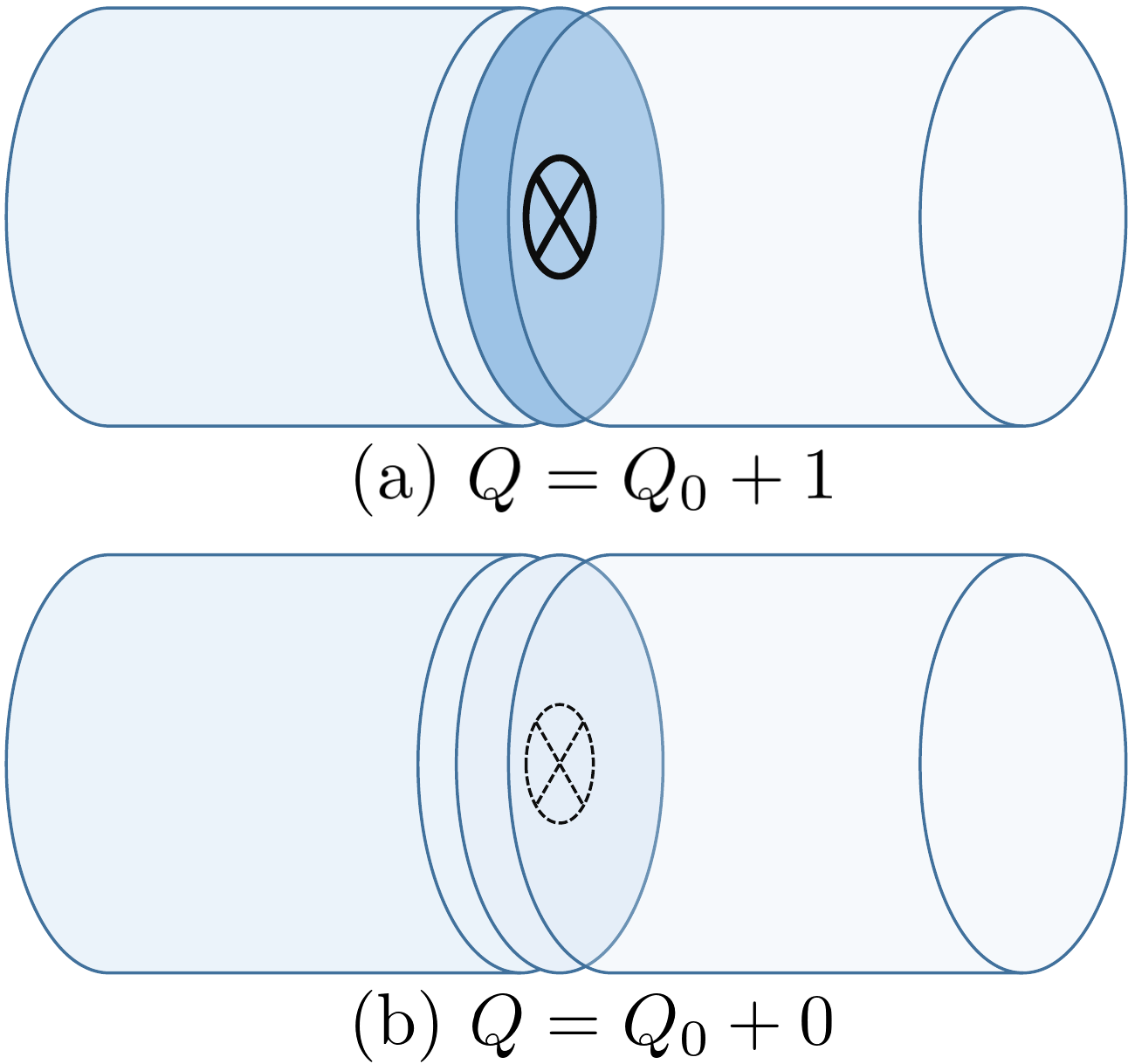}
\caption{In the space-like pump, a unit flux ``$\otimes$'' through the interface inserted in the vacuum induces nonzero additional charge if the interface supports nontrivial integer quantum Hall phase. S-transformed to the (temporal) pumps, such a flux can distinguish these two cycles.}
\label{Hall_pump}
\end{center}
\end{figure}

In this work, we will generalize the Thouless pump to higher order of response in arbitrary dimensions. 
In the viewpoint of space-like Thouless pump around Eq.~(\ref{afm}), 
the $U(1)$-charge accumulation there can be seen as alignment of spins within the finite regime exerted by the external magnetic field. 
Such a finite regime is zero-dimensional, which, in a large scale, can be seen as a point-like kink created by the external field. 
We can generalize this idea to higher dimensions, for example, three-dimensional space, the interface is a two-dimensional space. 
Although the charge jump of such an interface can still be discussed, we will later see that it vanishes once we do not perturb sufficiently many parameters to be spatially dependent. 
In this case, after S-transformed, the Thouless pump cannot distinguish topologically distinct cycles.
Nevertheless, integer quantum Hall conductance of this interface can obtain a nontrivial jump in its normal direction, e.g. we can smoothly insert an integer quantum Hall ($\sigma_\text{H}=1$) interface of a finite thickness in the vacuum as (a) in FIG.~(\ref{Hall_pump}). 
Such a smooth insertion can be effectively realized by a massive Dirac fermion with its mass term smoothly winding once by a space-dependent chiral transformation. 
Alternatively as (b) in FIG.~(\ref{Hall_pump}), 
we can insert a trivial phase $\sigma_\text{H}=0$ with the same total charge $Q_0$. 
These two interfaces cannot be adiabatically deformed to each other without closing the gap (of the whole system including the vacuum) although they accumulate the same number of charges $Q_0$ on the interface.
Therefore, after S-transformed to the temporal pump cycle, they represent two topologically distinct cycles and cannot be distinguished by the charge pump. 
It implies a generalization of the Thouless pump in higher dimensions, where the charge pump is generalized to higher-order $U(1)$ response pump (charge being the $0$-th order response), e.g. Hall-conductance pump, to distinguish those two cycles above. 

Furthermore, there is a reduction from higher-order response pumps to ($0$-th order) charge pumps by topologically nontrivial background fields as follows. 
In the $(3+1)$-dimensional example above, we can artificially insert a unit static flux through the interface.
Then, if the Hall conductances of the interfaces are differed by some integer, 
there will be the same quantized number of charge differences accumulated on the interface as in FIG.~(\ref{Hall_pump}). 
S-transformed to the temporal picture, there will be a nontrivial Thouless pump as a detector to distinguish these two cycles. 
Therefore, our generalization have observables through the Thouless pump with a nontrivial instanton background~\footnote{In general dimensions, the fluxes through intersecting surfaces result in nontrivial instantons}. 
Although motivated by the space-like picture above, we will still discuss the general pattern including original (time-like) pumps. 


The physical observable on the interfaces can also manifest the conflicting nature between parameter-space identification and symmetries, called $-1$-form anomaly~\cite{Cordova:2019aa,Cordova:2020aa} as follows. 
Gapped systems with a unique ground state are indistinguishable by the partition function in the low-energy limit ---
when we perform renormalization-group (RG) transformation on them, the asymptotic fixed-point models all have a single state. 
It is attempting to identify their parameters. 
However, such an identification can be potentially ill-defined once we couple the system to a background gauge field of a certain symmetry, e.g. $U(1)$-charge. 
To detect such an ambiguity, we can further couple the system to a background parameter gauge-field.
Such a parameter bundle, in the typical example of integer quantum Hall systems, will paste two distinct quantum Hall phases together at infinity. 
Nevertheless, as long as $U(1)$-symmetry is respected, the gap must be closed somewhere, such as on an interface, so that the Hall conductance can obtain a nonzero jump. 
Thus, the whole system cannot be gapped with a unique ground state any more. 
As a special situation, there can be chiral modes at some interface separating two distinct bulk quantum Hall phases. 
Such an inevitable ingappabilities signals an anomaly in the analog of the anomalous boundary modes on nontrivial symmetry-protected topological phases~\cite{Chen:2010aa,Chen:2013aa,Witten:2016ab,Witten:2019aa}. 
This anomaly is called $-1$-form symmetry anomaly since it is associated with the parameter space and the parameter change cannot be generated by quantum operators in real spacetime dimension(s). 
We will generalize the $-1$-form anomaly to parameter fields in a more general way that two parameter fields of distinct adiabatic phases to be identified necessarily depends on more spacetime coordinates than the example above. 

This article is organized as follows. 
The nonlinear parameter-gauge coupling action will be proposed in Sec.~(\ref{topology_parameter}), which implies a potentially nontrivial topology of parameter space. 
In Sec.~(\ref{Thouless_pump}), we will discuss the generalization of the Thouless pump to higher-order responses and their detection through the conventional charge pump by background fluxes. 
Finally, the concept of $-1$-form anomaly will be generalized in Sec.~(\ref{-1_form}) followed by various physical consequences related to interfaces.

\section{Nonlinear Parameter-gauge coupling}
\label{topology_parameter}
We consider spinless fermionic Hamiltonians respecting $U(1)$-charge symmetry. 
Let us assume that the gapped system is parameterized by a series of parameters $\{m_\alpha\}\in\mathcal{C}_D$ in the large-scale limit, where $\mathcal{C}_D$ is the set of parameters of $U(1)$-symmetric Hamiltonian with a unique ground state in $D=d+1$ spacetime dimension(s). 
In the continuum limit, the Lagrangian density takes the form as $\mathcal{L}(\{x_\mu\};\{m_\alpha\})$ with $\{x_\mu\}|_{\mu=0,\cdots,d}\equiv(t,x_1,x_2,\cdots,x_d)$. 
We can also perturb the parameters to be space-time dependent: 
\begin{eqnarray}
\mathcal{L}(\{x_\mu\};\{m_\alpha\})\Rightarrow\mathcal{L}(\{x_\mu\};\{\Phi_\alpha(x_\mu)\}), 
\end{eqnarray}
where $\Phi_\alpha(x_\mu)$ is the parameter field potentially depending on spacetime.

If the parameter perturbation is sufficiently smooth and locally insignificant,
the system is still gapped with a unique ground state and the ground state does not merge into higher energy, so all the matter fields can be integrated out. 
Generalizing the linear ``$A\wedge\cdots$'' coupling~\cite{Kapustin:2020aa,Kapustin:2020ab,Hsin:2020aa}, we can write down the gradient expansion of $\Phi_\alpha(x_\mu)$ of the topological actions of a general nonlinear coupling (NLC) among parameters and gauge field~\footnote{The theta-term type couplings, e.g. $(...)d\left[A(dA)^k\right]$, can be transformed to the Chern-Simons type by integration by part, which does not change the response of the form of $\left[-i\delta \ln\left(Z_\text{NLC}\right)/\delta A\right]$. }
in $D$-dimensional spacetime $S^1_{(0)}\times M$: 
\begin{eqnarray}
\label{effective}
Z_\text{NLC}[\{\Phi_\alpha(x)\},A]
=Z_0\exp\left(i\int_{S^1_{(0)}\times M}\sum_{k=0}^{\lfloor d/2\rfloor}\Phi^*[\lambda_k]\wedge\mathcal{L}^{(2k+1)}_\text{C-S}[A]\right), 
\end{eqnarray}
where $Z_0$ is independent on the gauge field and the Chern-Simons (C-S) density for general gauge field $\mathcal{A}$ is only defined in odd dimensions: 
\begin{eqnarray}
\mathcal{L}_\text{C-S}^{(1)}[\mathcal{A}]&\equiv&\text{Tr}\mathcal{A};\\
\mathcal{L}_\text{C-S}^{(3)}[\mathcal{A}]&\equiv&\frac{1}{4\pi}\text{Tr}\left(\mathcal{A}d\mathcal{A}+\frac{2}{3}\mathcal{A}^3\right);\cdots,
\end{eqnarray}
which, for $U(1)$ gauge fields, take the form as
\begin{eqnarray}
\mathcal{L}_\text{C-S}^{(2k+1)}=\frac{1}{(k+1)!}A\wedge\left(\frac{dA}{2\pi}\right)^k
\end{eqnarray}
The gradient expansion form $\lambda_k(m_\alpha)$ above is: 
\begin{eqnarray}
\lambda_k\equiv\sum_{\{\beta\}}\lambda_{k;\{\beta\}}(m_\alpha)dm_{\beta_1}\wedge dm_{\beta_2}\wedge\cdots\wedge dm_{\beta_{d-2k}}, 
\end{eqnarray}
which is pulled back by $\Phi^*$ to a differential form on the spacetime $S^1_{(0)}\times M$ from the parameter space $\mathcal{C}_D$: 
\begin{eqnarray}
\Phi^*[\lambda_k]=\sum_{\{\beta\},\{\gamma\}}\lambda_{k;\{\beta\}}[\Phi_\alpha(x)]\partial_{\gamma_1}\Phi_{\beta_1}dx^{\gamma^1}\wedge\cdots\wedge \partial_{\gamma_k}\Phi_{\beta_{d-2k}}dx^{\gamma_{d-2k}}, 
\end{eqnarray}
by which we can explicitly see the reason why the action (\ref{effective}) is topological --- a total antisymmetric tensor pops out so that the Lagrangian density is a Lorentz scalar without a prefactor $\sqrt{-g}$ where $g$ is the determinant of the metric. 
Moreover, torsion-free connection fields do not enter into the differentiations in the action due to the total asymmetric tensor, either. 

\section{Nontrivial topology of the parameter space $\mathcal{C}_D$}
Let us put the theory whose low-energy response is characterized by Eq.~(\ref{effective}) on a compact spin manifold $S^1_{(0)}\times M$. 
Then we do a gauge transformation $A\rightarrow A+d\theta$ on Eq.~(\ref{effective}), where $\theta(x)$ is a $2\pi$-periodic quantity. 
\begin{eqnarray}
&&Z_\text{NLC}[\{\Phi_\alpha(x)\},A+d\theta]\nonumber\\
&=&Z_\text{NLC}[\{\Phi_\alpha(x)\},A]\exp \left\{i\int_{S^1_{(0)}\times M}\sum_k\Phi^*[\lambda_k]\wedge d\theta\wedge\frac{1}{k!}\left(\frac{dA}{2\pi}\right)^k\right\}: 
\end{eqnarray}
which is restricted by the gauge-invariance: 
\begin{eqnarray}
\label{gauge-inv}
Z_\text{NLC}[\{\Phi_\alpha(x)\},A+d\theta]=Z_\text{NLC}[\{\Phi_\alpha(x)\},A]. 
\end{eqnarray}
A ``small'' gauge transformation --- $\theta(x)$ is single-valued and arbitrary --- gives the closedness condition
\begin{eqnarray}
d\Phi^*[\lambda_k]=\Phi^*[d\lambda_k]=0, 
\end{eqnarray}
by integration by part and that the differentiation commutes with the pull-back $d\Phi^*=\Phi^*d$. 
Since $\Phi$ can be perturbed locally from a constant function, we have the closeness for $\lambda_k$: 
\begin{eqnarray}
\label{closeness}
d\lambda_k=0. 
\end{eqnarray}

For large gauge transformations where $\theta(x)$ is multi-valued, we denote the Poincare dual of $d\theta/(2\pi)$ as $L_\theta$ which is $d$-cycle in $S^1_{(0)}\times M$. 
The effective action transforms as
\begin{eqnarray}
Z_\text{NLC}[\{\Phi_\alpha(x)\},A+d\theta]=Z_\text{NLC}[\{\Phi_\alpha(x)\},A]\exp\left\{i2\pi\oint_{L_\theta}\sum_k\Phi^*[\lambda_k]\wedge\frac{1}{k!}\left(\frac{dA}{2\pi}\right)^k\right\}.\nonumber\\
\end{eqnarray}
The gauge invariance (\ref{gauge-inv}) together with Eq.~(\ref{closeness}) and that $\theta(x)$ and $A$ are arbitrary implies
\begin{eqnarray}
\label{period}
\oint_{L_{d-2k}}\Phi^*[\lambda_k]=\oint_{\Phi_*(L_{d-2k})}\lambda_k
\in\mathbb{Z}, 
\end{eqnarray}
for each $k\in\{0,1,\cdots,\lfloor d/2\rfloor\}$ and arbitrary $(d-2k)$-cycle $L_{d-2k}\in Z_{d-2k}(S^1_{(0)}\times M,\mathbb{Z})$, and here $\Phi_*:Z_{d-2k}(S^1_{(0)}\times M,\mathbb{Z})\rightarrow Z_{d-2k}(\mathcal{C}_D,\mathbb{Z})$ is the push-forward. 

If the nontrivial period in Eq.~(\ref{period}) can be realized by some real system characterized by $\lambda_k(m_\alpha)$ with a closed brane $L_{d-2k}$ and $\Phi$, 
we can conclude that
\begin{eqnarray}
\mathcal{H}_{d-2k}(\mathcal{C}_D,\mathbb{Z})\supset\mathbb{Z}, 
\end{eqnarray}
where $\mathcal{H}_n(\mathcal{C}_D,\mathbb{Z})$ is the $n$-th homology of the parameter space $\mathcal{C}_D$. 
It is because, otherwise, the form $\lambda_k(m_\alpha)$ is exact, which sufficiently makes the integration in Eq.~(\ref{period}) vanish. 
Geometrically, a nontrivial integration in Eq.~(\ref{period}) means that there exists a gap-closing parameter in codimension $(D-2k)$, which can characterize the stability of criticality in real materials. 

\section{Generalizations of generalized Thouless pumps}
\label{Thouless_pump}
Let us take the following cycling: 
\begin{eqnarray}
\label{compactifiable}
\Phi^\text{P}_\alpha(\{x_\mu\}|_{\mu=0,\cdots,d})=\left\{\begin{array}{ll}\phi_\alpha(\{x_\mu\}|_{\mu=1,\cdots,d}),&t=x_0\rightarrow-\infty;\\
\phi_\alpha(\{x_\mu\}|_{\mu=1,\cdots,d}),&t=x_0\rightarrow+\infty,\end{array}\right. 
\end{eqnarray}
which means the system is periodic in time and we choose an adiabatic spacetime-dependent $\Phi^\text{P}_\alpha(\{x_\mu\})$ above so that the system remains gapped with a unique ground state at any time. 
For simplicity, we assume the space factorizing as $M=T^d$ so that our spacetime is compactified as $T^D$, but we will relax this condition later. 
We consider the order-$k$ response pump $\Sigma^\text{P}_{k;\mu_1\cdots\mu_{2k+1}}$ 
formally as the coefficient in the front of ``$dx^{\mu_1}\wedge\cdots\wedge dx^{\mu_{2k+1}}$'' component of $\mathcal{L}_\text{C-S}^{(2k+1)}$ in Eq.~(\ref{effective}) across a spacetime section perpendicular to $S^1_{(\mu_1)}\times\cdots\times S^1_{(\mu_{2k+1})}$: 
\begin{eqnarray}
\label{pump_0}
\Sigma^\text{P}_{k;\mu_1\cdots\mu_{2k+1}}\equiv\frac{1}{(d-2k)!}\epsilon^{\nu_1\cdots\nu_{d-2k}\mu_1\cdots\mu_{2k+1}}\oint_{\Phi^\text{P}_*\left[S^1_{(\nu_1)}\times\cdots\times S^1_{(\nu_{d-2k})}\right]}\lambda_k\in\mathbb{Z}, 
\end{eqnarray}
where $\{\nu\}$ is summed implicitly. 
Nontrivial contributions to the integration above result from any oriented spacetime section $S^1_{(\nu_1)}\times\cdots\times S^1_{(\nu_{d-2k})}$ perpendicular to the $x_{\mu_1}$-$\cdots x_{\mu_{2k+1}}$-axes, and such an integration is well-defined since any pair of these sections are cobordant and $\Phi^{\text{P}*}[\lambda_k]$ is closed. 
Actually, we can have a more general setting where $S^1_{(\nu_1)}\times\cdots\times S^1_{(\nu_{d-2k})}$ is replaced by any other spin manifold. 
In a similar sense, even the temporal pump can be made space-like by switching $x_0$ from $t$ to any other spatial component. 
Since $\lambda_k$ is locally a $(d-2k)$-form, we need at least $(d-2k)$ parameters to construct $\Phi^\text{P}_\alpha$ so that the period (\ref{pump_0}) can be nontrivial. 
In addition, if $d-2k=0$, then the period above vanishes.
Physically, it implies the fact that the highest response in even-dimensional space is constant as long as the gap does not close. 

We calculate the net charge flow through the section in one cycle $t\in(-\infty,+\infty)$ perpendicular to the $x_i$-direction as: 
\begin{eqnarray}
\label{pump_1}
Q^\text{P}_i&=&\oint_{S^1_{(0)}\times\cdots{S}^1_{(i-1)}\times\hat{S}^1_{(i)}\times{S}^1_{(i+1)}\cdots S^1_{(d)}}\frac{\delta}{i\delta A_i} \ln(Z_\text{NLC})\nonumber\\
&=&(-1)^{d-i}\sum_k\oint_{S^1_{(0)}\times\cdots{S}^1_{(i-1)}\times\hat{S}^1_{(i)}\times{S}^1_{(i+1)}\cdots S^1_{(d)}}\Phi^{\text{P}*}[\lambda_k]\wedge\imath_i^*\left[\frac{1}{k!}\left(\frac{dA}{2\pi}\right)^k\right]\nonumber\\
&=&(-1)^{d-i}\sum_k\oint_{\Phi^\text{P}_*\left[M^{(i)}_{k}\right]}\lambda_k, 
\end{eqnarray}
where $\imath_i$ is the inclusion of the oriented $S^1_{(0)}\times\cdots{S}^1_{(i-1)}\times\hat{S}^1_{(i)}\times{S}^1_{(i+1)}\cdots S^1_{(d)}$ into the full spacetime,
and $M^{(i)}_k$ is the Poincare dual to the pullback $\imath_i^*[(dA/2\pi)^k/k!]$. 
Since the integration in the second line above is done on a closed form, it is independent of the choice $S^1_{(0)}\times\cdots{S}^1_{(i-1)}\times\hat{S}^1_{(i)}\times{S}^1_{(i+1)}\cdots S^1_{(d)}$ within the same homology. 
It is related to the terms in Eq.~(\ref{pump_1}) as follows.

\subsection{Reduction to $D>0,k=0$: Thouless pumps in high dimensions}
The Thouless pump (T-P) in high dimensions~\cite{Kapustin:2020ab} implies that a spacetime-dependent (depending on $d$ coordinates) parameter, will change by an integer number the total charges in the section perpendicular to the hypersurface spanned by those $d$ coordinates. 
It can be deduced from our result in Eq.~(\ref{pump_1}) as a special case of $k=0$ with $x_i$ the $i$-th spatial component and $dA=0$: 
\begin{eqnarray}
Q^\text{P}_i=\Sigma^\text{P}_{0;i}
=(-1)^{d-i}\oint_{\Phi^\text{P}_*\left[S^1_{(0)}\times\cdots\times\hat{S}^1_{(i)}\times\cdots S^1_{(d)}\right]}\lambda_0\in\mathbb{Z}, 
\end{eqnarray}
where $\hat{}$ means a deletion. 
Since $\lambda_0$ is locally a $d$-form, 
it is necessary to have at least $d$ parameters in order that the period above can be nontrivial. 

\subsection{Detection related to higher responses: nontrivial background instantons}
As mentioned in the Introduction part,
we can measure the higher-order response pump by adiabatic flux insertions~\cite{Oshikawa:2000aa} or static flux configuration, or formally, winding surfaces of spacetime manifold around monopoles --- instantons. 
In the presence of nontrivial background instanton,
we obtain that the total charge pump along the $x_i$-direction is determined by higher-order responses: 
\begin{eqnarray}
Q_i^\text{P}=\Sigma_{0;i}^\text{P}+\sum_{k=1}^{\lfloor d/2\rfloor}\frac{1}{(2k)!}\mathcal{P}_{\mu_1\cdots\mu_{2k}}\Sigma^\text{P}_{k;\mu_1\cdots\mu_{2k}i}, 
\end{eqnarray}
where the spacetime indices $\{\mu\}$ are summed implicitly in advance to the summation of $k$, and $\mathcal{P}_{\mu_1\cdots\mu_{2k}}$ is the instanton number: 
\begin{eqnarray}
\mathcal{P}_{\mu_1\cdots\mu_{2k}}\equiv\oint_{S^1_{(\mu_1)}\times\cdots\times S^1_{(\mu_{2k})}}\frac{1}{k!}\left(\frac{dA}{2\pi}\right)^k\in\mathbb{Z}. 
\end{eqnarray}
If we switch $i$-th direction above to be temporal,
we would have obtained the corresponding spatial-like Thouless pump as the spin example in the Introduction. 

\subsubsection{Example: adiabatic flux insertion in $D=3+1$ --- a Laughlin-type argument}
Let us adiabatically insert a flux threading the $x_1$-loop: 
\begin{eqnarray}
\oint_{S^1_{(0)}\times S^1_{(1)}}dA=2\pi N_\text{flux}, 
\end{eqnarray}
which is adiabatic enough so that the initial many-body state remains staying at the ground state, and $A_2=A_3=0$. 
Then we can obtain: 
\begin{eqnarray}
Q_{i}^\text{P}=\Sigma^\text{P}_{0;i}+N_\text{flux}\Sigma^\text{P}_{1;01i},\,\,\,i\in\{2,3\}. 
\end{eqnarray}
In addition to the first term which is Thouless-pumped charge, the second term is understood as, by the Laughlin argument~\cite{Laughlin:1981aa}, the transverse charge flow induced by the electric field $\partial_0A_1$ due to the Hall conductance of a spatial slice spanned by $S^1_{(1)}\times S^1_{(i)}$. 

\subsubsection{Example: static flux configuration in $D=3+1$ --- a magneto-electric effect}
We add a static background flux through the plane $S^1_{(1)}\times S^1_{(2)}$: 
\begin{eqnarray}
\oint_{S^1_{(1)}\times S^1_{(2)}}\left[\triangledown\times\vec{A}\right]_{x_3}=\oint_{S^1_{(1)}\times S^1_{(2)}}dA=2\pi N_\text{flux}, 
\end{eqnarray}
with vanishing $A_0=A_3=0$. Then
\begin{eqnarray}
Q_3^\text{P}=\Sigma^\text{P}_{0;3}+N_\text{flux}\Sigma^\text{P}_{1;123}. 
\end{eqnarray}
The first term $\Sigma^\text{P}_{0;3}$ is the net Thouless-pumped charge flowing along $x_3$-direction in the absence of background fluxes. 
As a special case in the three-dimensional topological insulator, 
the second term exactly reflects the magneto-electric effect~\cite{Qi:2008aa,Qi:2011aa},
where $\Sigma_{1;123}^\text{P}$ is the net jump of theta-term induces a change of polarization in the presence of a static flux $2\pi N_\text{flux}$ through the plane spanned by $S^1_{(1)}\times S^1_{(2)}$.

\section{Generalized $-1$-form anomalies}
\label{-1_form}
Since the $D$-dimensional systems parametrized by $\mathcal{C}_D$ are defined to be gapped, 
their Hilbert spaces, at the low-energy limit, consisting of one single state cannot be distinguished by the partition function which is trivially $1$. 
It is attempting to identify their RG fixed point parameters which are functions $M\rightarrow\mathcal{C}_D$ and called parameter fields. 
However, they might be distinguishable once coupled with a background gauge field of global symmetries. 
Therefore, the certain global symmetries obstruct such an identification. 
As we will see below, this obstruction has the physical consequence as inevitable gaplessness or spectrum degeneracy of spatial interfaces, or gap closing during a time-dependent procedure. 
Actually, we have already seen such phenomena in symmetry-respected interfaces between topological insulators or other invertible phases, e.g. $U(1)$-respected interfaces carrying chiral modes between two distinct integer quantum Hall phases. 
We will first generalize $-1$-form anomaly and then the physical consequence of anomalous systems will be presented followed by various examples. 

\subsection{Obstructions by parameter bundles}
We can define a $-1_{[n]}$-form anomaly for $n=d\mod2$ by the identification of 
two RG-fixed parameter functions
$\mathcal{T}_\alpha\sim\mathcal{K}_\alpha:M\rightarrow \mathcal{C}_D$
satisfying: 
\begin{eqnarray}
\label{identification}
\oint_{U^{(n)}}\mathcal{T}^*\left[\lambda_{(d-n)/2}\right]\neq\oint_{V^{(n)}}\mathcal{K}^*\left[\lambda_{(d-n)/2}\right], 
\end{eqnarray}
where the integration is performed on some $n$-dimensional closed subspaces $U^{(n)}$ and $V^{(n)}$ of $M$, which are in the same homological class of $\mathcal{H}_n(M,\mathbb{Z})$. 
The temporal component $S^1_{(0)}$ is taken as the one-point compactification of $\mathbb{R}^1_{(0)}$. 
The anomaly of such an identification can be detected by the following parameter spacetime bundle: 
\begin{eqnarray}
\label{boundary}
\Phi=\left\{\begin{array}{ll}\mathcal{T}\circ \pi_M,&x_{0}\rightarrow-\infty;\\\mathcal{K}\circ\pi_M,&x_{0}\rightarrow+\infty,\end{array}\right. 
\end{eqnarray}
where $\pi_M$ is the natural projections $S_{(0)}^1\times M\rightarrow M$. 
Due to the identification $\mathcal{T}\sim\mathcal{K}$, the bundle (\ref{boundary}) is well-defined at $x_0=\infty$ (or the one-point compactification $\{x_0\rightarrow-\infty\}\sim\{x_0\rightarrow+\infty\}$ is well-defined). 
Such a construction is similar to mapping tori~\cite{Witten:1985aa,Witten:2016ab}, which is applied to detect conventional anomalies, while the mapping tori is, in our situation, extended along one of the physical dimensions, e.g. $x_0$-component, rather than an extra dimension. 
However, the gauge-invariance condition
\begin{eqnarray}
\label{gauge_inv_condition}
d\Phi^*\left[\lambda_{(d-n)/2}\right]=0
\end{eqnarray}
requires, by $\{-\infty\}\times V^{(n)}$ and $\{+\infty\}\times U^{(n)}$ being in the same homology class, that
\begin{eqnarray}
\oint_{\{-\infty\}\times V^{(n)}}\Phi^*\left[\lambda_{(d-n)/2}\right]=\oint_{\{+\infty\}\times U^{(n)}}\Phi^*\left[\lambda_{(d-n)/2}\right], 
\end{eqnarray}
which contradicts with Eq.~(\ref{identification}). 
Physically, it means that the following three aspects are contradicting: the global $U(1)$ symmetry, being gapped with a unique ground state~\footnote{Being gapped here means the system is gapped at any time slice of the spacetime $S^1_{(0)}\times M$.}, and the parameter identification ``symmetry'' $\mathcal{T}\sim\mathcal{K}$. 
We denote such a mixed anomaly by $-1_{[n]}$-form anomaly. 
In the following, we will discuss its dynamical consequences. 

\subsection{Physical consequences: gap closing during a time-dependent procedure}
Let us first discuss the consequence of the contradiction between the parameter bundle (\ref{boundary}) and the requirement (\ref{gauge_inv_condition}). 
It should be noted that such a condition (\ref{gauge_inv_condition}) is derived from the gauge invariance of the effective response action (\ref{effective}) which is already presumed to be gapped with a unique ground state at any time slice. 
Thus Eq.~(\ref{gauge_inv_condition}) actually implies both the gauge invariance and being gapped non-degenerately. 
Alternatively, Eq.~(\ref{boundary}) can describe the asymptotic behavior of a realistic time-dependent procedure supported on the spacetime $\mathbb{R}^1_{(0)}\times M$. 
The $-1_{[n]}$-form anomaly above means that the gap of the system must be closed at some time $t_0\in(-\infty,+\infty)$ no matter how does the system evolve meanwhile. 
\begin{itemize}
\item {Example: spinless fermions with $d=0$ and $n=0$}

For a single gapped spinless fermion $\psi$ with a unique ground state in $(0+1)$ dimension, 
\begin{eqnarray}
\mathcal{L}_\text{fermion;(0,0)}(m_\psi)&=&i{\psi}^\dagger\left(\partial_t-i{A}_0\right)\psi-m_\psi{\psi}^\dagger\psi+\mathcal{L}_\text{reg};\nonumber\\
\mathcal{L}_\text{reg}&=&i{\chi}^\dagger\left(\partial_t-i{A}_0\right)\chi+\mu_\chi{\chi}^\dagger\chi, 
\end{eqnarray}
where $\chi$ is a dynamical bosonic spinor as a regulator $\mu_\chi\rightarrow+\infty$.
The effective response action takes the form as
\begin{eqnarray}
\mathcal{S}_\text{fermion;0}=\int dt\,\,\Theta(m_\psi)A_0(t), 
\end{eqnarray}
where $\Theta(m_\psi)$ is the step function nonvanishing $1$ only when its argument is positive. 
Additionally, $\mathcal{H}_0(\mathcal{C}_1,\mathbb{Z})\supset\mathbb{Z}\oplus\mathbb{Z}$ where the first factor corresponds to $m_\psi<0$ and the second one to $m_\psi>0$ and then a gapless point, e.g. $m_\psi=0$, must occur in co-dimension at least $1$ in the full parameter space.
 
Such a system has a $-1_{[0]}$-form anomaly associated with the identification $\mathcal{T}_{m_\psi}\sim\mathcal{K}_{m_\psi}$ where $\mathcal{T}_{m_\psi}(\vec{x})=-\mu_\chi=-\mathcal{K}_{m_\psi}(\vec{x})$ and this anomaly is detected by the bundle
\begin{eqnarray}
\Phi_{m_\psi}=\left\{\begin{array}{ll}\mathcal{T}_{m_\psi}\circ \pi_M=-\mu_\chi,&x_{0}\rightarrow-\infty;\\\mathcal{K}_{m_\psi}\circ\pi_M=+\mu_\chi,&x_{0}\rightarrow+\infty,\end{array}\right. 
\end{eqnarray}
of the form of Eq.~(\ref{boundary}). 

On the other hand, if we see this bundle as the asymptotic condition for a time-dependent procedure supported on $\mathbb{R}^1_{(0)}\times M$, 
the gap must be closed at some intermediate time point as long as $U(1)$ symmetry is respected by the Hamiltonian during $x_0\in(-\infty,+\infty)$. 
Actually, this gap closing can be clearly seen once $\Theta(m_\psi)$ here is noticed as $U(1)$ charge, whose ground-state eigenvalue cannot change unless the gap closes. 
Thus the $-1_{[0]}$-form anomaly of a single spinless fermion in $(0+1)$ dimension associated with the identification above signals an inevitable gap closing during a related time-dependent procedure. 
\end{itemize}
\subsection{Physical consequences: ingappabilities within a spatial interface}
For $d-n\neq0$, 
it is also consistent to exist the following $-1_{[n]}$-form anomaly associated with the identification $\mathcal{T}\sim\mathcal{K}:M\equiv S_{(\nu)}^1\times M_{d-1}\rightarrow\mathcal{C}_D$, where $\mathcal{T}$ and $\mathcal{K}$ does not depend on $x_\nu\in S_{(\nu)}^1$, satisfying
\begin{eqnarray}
\label{identification_space}
\oint_{u^{(n)}}\mathcal{T}^*\left[\lambda_{(d-n)/2}\right]\neq\oint_{v^{(n)}}\mathcal{K}^*\left[\lambda_{(d-n)/2}\right], 
\end{eqnarray}
for some $u^{(n)}$ and $v^{(n)}$ in the same homology class of $\mathcal{H}_n(M_{d-1},\mathbb{Z})\subset\mathcal{H}_n(M,\mathbb{Z})$ and $S^1_{(\nu)}$ is a special spatial component parametrized by $x_\nu$ and one-point compactification of $\mathbb{R}^1_{\nu}$. Once the identification $\mathcal{T}\sim\mathcal{K}$ is made, we can consider the following spacetime parameter bundle: 
\begin{eqnarray}
\label{boundary_space}
\Phi=\left\{\begin{array}{ll}\mathcal{T}\circ \pi_{M},&x_{\nu}\rightarrow-\infty;\\\mathcal{K}\circ\pi_{M},&x_{\nu}\rightarrow+\infty,\end{array}\right. 
\end{eqnarray}
which is well-defined since neither $\mathcal{T}$ nor $\mathcal{K}$ is $x_\nu$-dependent. 

Similarly to the time-like case before, such a bundle (\ref{boundary_space}) contradicts with the gauge invariance (\ref{gauge_inv_condition}) which already presumes the system is gapped non-degenerately. 
Therefore, if we have a space-like (time-independent) interface along $\mathbb{R}^1_{(\nu)}$ supported on $S^1_{(0)}\times\mathbb{R}^1_{(\nu)}\times M_{d-1}$, whose asymptotic form obeys Eq.~(\ref{boundary_space}), the whole system cannot be gapped, e.g. there can be a localized massless mode along a $\tilde{x}_\nu$-slice for some $\tilde{x}_\nu\in\mathbb{R}^1_{(\nu)}$. 
\begin{itemize}
\item {Example: $(2+1)$-dimensional fermion with $d=2$ and $n=0$}

Let us consider the fermionic model in $(2+1)$ dimensions: 
\begin{eqnarray}
\mathcal{L}_\text{fermion;(2,1)}(m_\psi)&=&i\bar{\psi}\left(\slashed{\partial}-i\slashed{A}\right)\psi-m_\psi\bar{\psi}\psi+i\bar{\chi}\left(\slashed{\partial}-i\slashed{A}\right)\chi+\mu_\chi\bar{\chi}\chi, 
\end{eqnarray}

The effective action can be obtained after the matter fields $\psi$ and $\chi$ are integrated out: 
\begin{eqnarray}
\mathcal{S}_\text{fermion;(2,1)}(m_\psi)=\int \Theta(m_\psi)\frac{1}{4\pi}AdA. 
\end{eqnarray}
It means $\mathcal{H}_0(\mathcal{C}_3,\mathbb{Z})\supset\mathbb{Z}\oplus\mathbb{Z}$ of which the first factor results from $m_\psi<0$ and the second one from $m_\psi>0$. 
Thus a gapless point, e.g. $m_\psi=0$, must exist and occur in co-dimension at least $1$. 

Furthermore, we have a $-1_{[0]}$-form associated with the identification of RG-fixed points $\{m_\psi=-\mu_\chi\}\sim\{m_\psi=+\mu_\chi\}$. 
In addition to the time-like detection by the bundle (\ref{boundary}), 
we can alternatively apply the parameter bundle of the form of Eq.~(\ref{boundary_space}): 
\begin{eqnarray}
\Phi_{m_\psi}=\left\{\begin{array}{ll}\mathcal{T}_{m_\psi}\circ \pi_M=-\mu_\chi,&x_1\rightarrow-\infty;\\\mathcal{K}_{m_\psi}\circ\pi_M=+\mu_\chi,&x_1\rightarrow+\infty,\end{array}\right. 
\end{eqnarray}
with $M=S^1_{(1)}\times S^1_{(2)}$, to detect the anomaly. 

The bundle above can also describes a realistic space-like interface along $\mathbb{R}_{(1)}^1$ supported on $S^1_{(0)}\times\mathbb{R}_{(1)}^1\times S^1_{(2)}$ and the anomaly implies that such an interface must have degenerate ground states or be gapless. 
Indeed, it reproduces the fact that two distinct quantum Hall phases $\sigma_\text{H}=\Theta(m_\psi=-\mu_\chi)=0$ and $\sigma_\text{H}=\Theta(m_\psi=+\mu_\chi)=1$ cannot be spatially adiabatically connected without a gap closing as long as the Hamiltonian respects $U(1)$ symmetry. 
One possible choice is the emergence of localized massless chiral fermion on the interface between two phases. 

\item {Example: (3+1)-dimensional fermion with $d=3$ and $n=1$}

We can also consider the following $(3+1)$-dimensional $U(1)$-symmetric Dirac fermion: 
\begin{eqnarray}
\mathcal{L}_\text{fermion;(3,1)}(\varphi)&=&i\bar{\psi}\left(\slashed{\partial}-i\slashed{A}\right)\psi+m_\psi\bar{\psi}\exp(i\gamma^5\varphi)\psi+\mathcal{L}_\text{reg}[\bar{\chi},\chi,A]\nonumber\\
\mathcal{L}_\text{reg}&=&i\bar{\chi}\left(\slashed{\partial}-i\slashed{A}\right)\chi+\mu_\chi\bar{\chi}\chi, 
\end{eqnarray}
where $\gamma^5\equiv\gamma^0\gamma^1\gamma^2\gamma^3$ and $\varphi$ is a real parameter. 
If we take $m_\psi>0$, the effective response action takes the form as
\begin{eqnarray}
\mathcal{S}_\text{fermion;(3,1)}=\int \Phi_\varphi^*\left[\frac{d\varphi}{2\pi}\right]\wedge \frac{1}{4\pi}AdA, 
\end{eqnarray}
after the matter fields $\psi$ and $\chi$ are integrated out. 
The gauge invariance condition (\ref{gauge_inv_condition}) requires that $\varphi$ is a $2\pi$-periodic parameter 
which is obvious in $\mathcal{L}_\text{fermion;(3,1)}$ above. Furthermore, $d\varphi/2\pi$ is in a nontrivial closed form, which means a gapless point, e.g. $m_\psi=0$, occurs in co-dimension, at least, $2$, in the full parameter space. 

Considering the theory on the spacetime $S^1_{(0)}\times M=S^1_{(0)}\times S^1_{(1)}\times S^1_{(2)}\times S^1_{(3)}$ with $S^1_{(1)}$ one-point compactification of $\mathbb{R}_{(1)}^1$, we can obtain a $-1_{[1]}$-form anomaly associated with the identification $\mathcal{T}\sim\mathcal{K}$ where $\mathcal{T}_\varphi(\vec{x})=2\pi R_\mathcal{T}x_2/L_2+2\pi N_\mathcal{T}x_3/L_3$ and $\mathcal{K}_\varphi(\vec{x})=2\pi R_\mathcal{K}x_2/L_2+2\pi N_\mathcal{K}x_3/L_3$ with distinct integer pairs $(R_\mathcal{T},N_\mathcal{T})\neq(R_\mathcal{K},N_\mathcal{K})$ and $L_3$ the length of $S^1_{(3)}$. 
Such an anomaly can be detected by the space-like parameter bundle of the form of Eq.~(\ref{boundary_space}): 
\begin{eqnarray}
\label{boundary_space_example}
\Phi_\varphi=\left\{\begin{array}{ll}\mathcal{T}_\varphi\circ \pi_{M}=2\pi R_\mathcal{T}x_2/L_2+2\pi N_\mathcal{T}x_3/L_3,&x_1\rightarrow-\infty;\\\mathcal{K}_\varphi\circ\pi_{M}=2\pi R_\mathcal{K}x_2/L_2+2\pi N_\mathcal{K}x_3/L_3,&x_1\rightarrow+\infty,\end{array}\right. 
\end{eqnarray}
which contradicts with the gauge invariance condition $d\Phi_\varphi^*[\varphi]=0$ since
\begin{eqnarray}
\left(\begin{array}{l}R_\mathcal{T}\\ N_\mathcal{T}\end{array}\right)=\left(\begin{array}{l}\oint_{S^1_{(2)}\subset S^1_{(2)}\times S^1_{(3)}}\\\oint_{S^1_{(3)}\subset S^1_{(2)}\times S^1_{(3)}}\end{array}\right)\mathcal{T}_\varphi^*\left[\frac{d\varphi}{2\pi}\right]\neq\left(\begin{array}{l}\oint_{S^1_{(2)}\subset S^1_{(2)}\times S^1_{(3)}}\\\oint_{S^1_{(3)}\subset S^1_{(2)}\times S^1_{(3)}}\end{array}\right)\mathcal{K}_\varphi^*\left[\frac{d\varphi}{2\pi}\right]=\left(\begin{array}{l}R_\mathcal{K}\\ N_\mathcal{K}\end{array}\right). \nonumber\\
\end{eqnarray}
The bundle (\ref{boundary_space_example}) above can also describe (the asymptotic form of) a space-like interface along $\mathbb{R}^1_{(1)}$ supported on $S^1_{(0)}\times\mathbb{R}^1_{(1)}\times S^1_{(2)}\times S^1_{(3)}$.
The $-1_{[1]}$-form anomaly here signals that this interface must be degenerate or gapless. 
\begin{figure}[h]
\begin{center}
\includegraphics[width=9.5cm,pagebox=cropbox,clip]{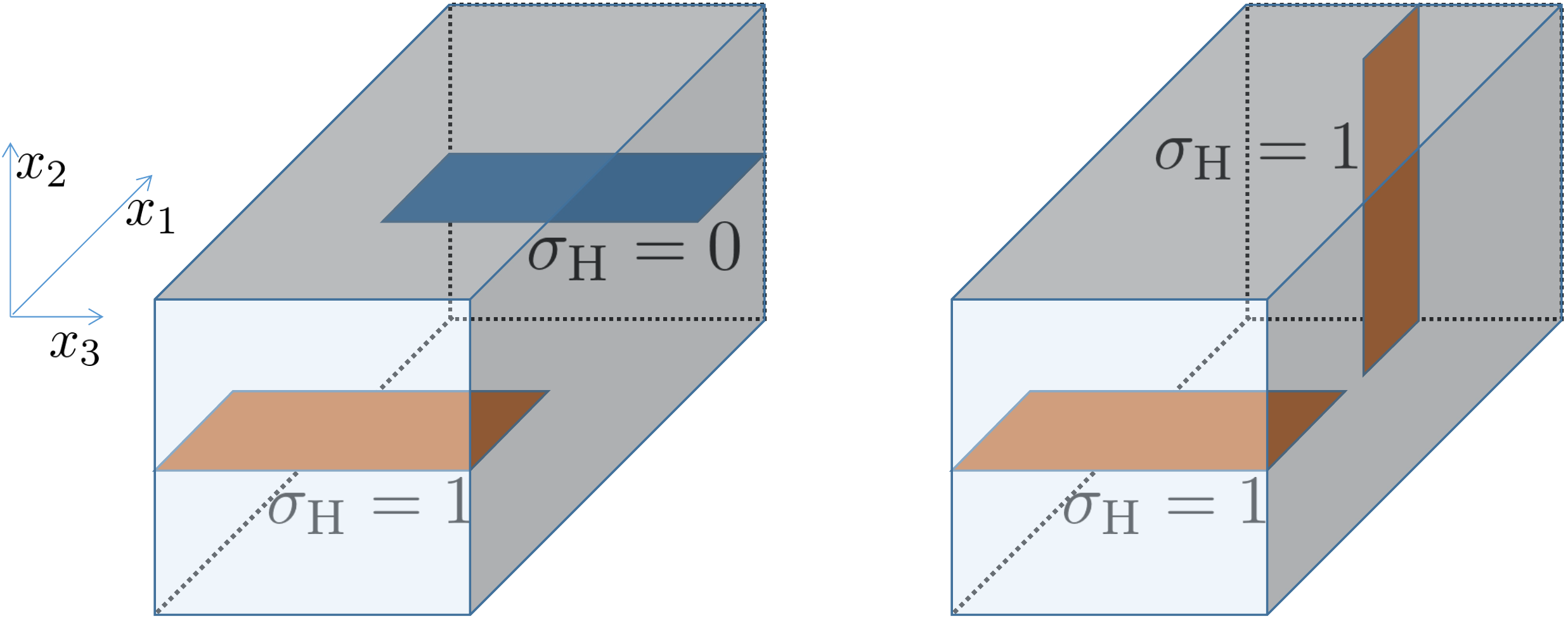}
\caption{When the $\varphi$-angle rotates ``locally'' as Eq.~(\ref{locally}), the interface with the asymptotic condition $(R_\mathcal{T},N_\mathcal{T})=(1,0)$ and $(R_\mathcal{K},N_\mathcal{K})=(0,0)$ is depicted on the left and the interface with the asymptotic condition $(R_\mathcal{T},N_\mathcal{T})=(1,0)$ and $(R_\mathcal{K},N_\mathcal{K})=(0,1)$ is sketched on the right. 
Here the $x_2$-$x_3$ plane is a torus $T^2$ and the direction $x_1$ is the non-compact component. }
\label{Examples}
\end{center}
\end{figure}
Physically, such an interface setting, whose asymptotic behavior is described by (\ref{boundary_space_example}), can be seen as two pairs of perpendicular Hall planes if we make the $\varphi$-rotation in (\ref{boundary_space_example}) more ``local'', e.g. by the substitutions in Eq.~(\ref{boundary_space_example}) as: 
\begin{eqnarray}
\label{locally}
2\pi R_\mathcal{T,K}x_2/L_2\mapsto 2\pi R_\mathcal{T,K}\Theta(x_2)\text{  if  }x_2\notin[-l_2/2,+l_2/2];\nonumber\\
2\pi N_\mathcal{T,K}x_3/L_3\mapsto 2\pi N_\mathcal{T,K}\Theta(x_3)\text{  if  }x_3\notin[-l_3/2,+l_3/2],
\end{eqnarray}
with $l_{2,3}\ll L_{2,3}$. 
After this modification, as a special case, $(R_\mathcal{T},N_\mathcal{T})=(1,0)$ and $(R_\mathcal{K},N_\mathcal{K})=(0,0)$ means that we have a quantum Hall piece with $\sigma_\text{H}=1$ on some $x_1$-$x_3$ plane at the infinity $x_1=-\infty$ while no nontrivial quantum Hall piece at the infinity $x_1=+\infty$ as the left part in FIG.~(\ref{Examples}). 
Thus it is impossible to adiabatically connect both sides in the middle spatial regime $x_1\in(-\infty,+\infty)$.  
Furthermore, we can also have another different special interface setting $(R_\mathcal{T},N_\mathcal{T})=(1,0)$ and $(R_\mathcal{K},N_\mathcal{K})=(0,1)$ as the right part in FIG.~(\ref{Examples}). 
Although both the three-dimensional spatial regimes around $x_1=-\infty$ and $x_1=+\infty$ support $\sigma_\text{H}=1$ quantum Hall pieces, they still cannot be adiabatically connected in the whole space. 
It is because their supports, $x_1$-$x_3$ plane and $x_1$-$x_2$ plane, cannot be continuously pasted in the intermediate areas since $S^1_{(3)}\subset M$ and $S^1_{(2)}\subset M$ are homologically inequivalent. 
On the other hand, 
if we replace the torus $S^1_{(2)}\times S^1_{(3)}$ by a topologically trivial disk $[-L_2/2,+L_2/2]\times[-L_3/2,+L_3/2]=I^1_{(2)}\times I^1_{(3)}$,
the two asymptotic sides can be adiabatically connected, e.g. by spiral of the Hall piece by $\pi/2$-angle. 
\end{itemize}
In a short summary, each $-1_{[n]}$-form anomaly associated with an identification of two RG-fixed parameter fields can be detected by a parameter bundle on a closed spacetime manifold. 
Such a bundle can be also alternatively seen as a time-dependent procedure or a time-independent spatial interface supported by a (non-compact) spacetime manifold, where the system inevitably undergoes a gap closing during the procedure or within the intermediate spatial regime, respectively. 

\section{Conclusions}
By a nonlinear parameter-gauge topological response theory, we investigated the nontrivial topology of the parameter space of general $U(1)$-symmetric fermionic non-degenerately gapped system and its consequences on the transport properties in arbitrary dimensions. 
Such nontrivial topology can impose quantization constraints on the charge transport in the presence of background fluxes or, more generally, instantons in general dimensions.
These transport properties are related to a parameter-coupling quantum anomaly where the parameter dependence on spacetime generalizes $-1$-form anomalies. 
This anomaly imposes non-perturbative ingappabilities of various types of spatial interfaces or time-dependent system evolution. 
In this work, we use spin manifolds as the underlying spacetime to detect the topology of the parameter space and the generalization to arbitrary spin$^c$ manifolds would be of future interest. 

\acknowledgements
The author was supported by JSPS fellowship.
This work was supported by JSPS KAKENHI Grant Nos. JP19J13783. 
A part of the present work was performed at Kavli Institute for Theoretical Physics, University of California at Santa Barbara, supported by
US National Science Foundation Grant No. NSF PHY-1748958.

\sloppy
%

\end{document}